\documentclass[twocolumn,aps,showpacs,pra,amsmath]{revtex4}

\usepackage{graphics}
\usepackage[pdftex]{graphicx}
\usepackage[pdftex]{color}

\begin{document}

\title{Large peak-to-valley ratio of negative-differential-conductance \\ in graphene p-n junctions}

\author{V. Hung Nguyen$^{1,2}$\footnote{E-mail:
viet-hung.nguyen@u-psud.fr}, A. Bournel$^{1}$ and P. Dollfus$^{1}$}

\address{$^{1}$Institut d'Electronique Fondamentale, UMR8622, CNRS, Univ. Paris Sud, 91405 Orsay, France \\
$^{2}$Center for Computational Physics, Institute of Physics, VAST, P.O. Box 429 Bo Ho, Hanoi 10000, Vietnam}

\date{\today}
\begin{abstract}
  We investigate the transport characteristics of monolayer graphene p-n junctions by means of the non-equilibrium Green's function technique. It is shown that thanks to the high interband tunneling of chiral fermions and to a finite bandgap opening when the inversion symmetry of graphene plane is broken, a strong negative-differential-conductance behavior with peak-to-valley ratio as large as a few tens can be achieved even at room temperature. The dependence of this behavior on the device parameters such as the Fermi energy, the barrier height, and the transition length is then discussed.
\end{abstract}

 \maketitle

\section{Introduction}
Since the discovery of the Esaki tunnel diode \cite{esak58}, the effect of negative-differential-conductance (NDC) has given rise to an intense research activity from fundamental aspects of transport to possible applications including oscillator, frequency multiplier, memory, fast switching, etc. \cite{mizu95}. Motivated by the recent development of graphene nanoelectronics \cite{neto09,schw10,lin010}, this effect has been also investigated and discussed in some nanostructures based on monolayer graphene \cite{vndo08,chau09}, bilayer graphene \cite{hung09}, and graphene nanoribbons \cite{ren009,lian10,vndo10}. However, these studies have shown that due to their zero-bandgap, the peak-to-valley ratio (PVR) of NDC in 2D-graphene structures is relatively small, while, though it can be improved significantly in narrow graphene nanoribbons, the results obtained are severely affected by increasing the ribbon's width and/or, especially, by the edge defects \cite{lian10,vndo10}.

A bandgap opening may be a key-step to improve the operation of graphene devices. Beyond the technique which consists in patterning a graphene sheet into nanoribbons \cite{mhan07}, some recent works have suggested that a bandgap can open when the inversion symmetry of the graphene plane is broken, e. g., by the interaction with the substrate \cite{zhou07,vita08,ende10}, the patterned hydrogen adsorption \cite{balo10}, the adsorption of water molecules \cite{yava10}, or by the controlled structural modification \cite{echt08}. In particular, the experiment reported in \cite{zhou07} has shown that graphene epitaxially grown on SiC can exhibit a bandgap up to $260$ meV. Though relatively small compared to that in conventional semiconductors, it is about ten times greater than the thermal energy at room temperature, which has stimulated further investigations of graphene-based transistors \cite{lian07,kedz08,mich10}. In this work, our aim is to look for possibilities of achieving a strong NDC behavior in monolayer graphene p-n junctions. This expectation comes from the fact that the appearance of a finite bandgap can result in a low valley current, while, the high interband tunneling of chiral fermions in graphene may lead to a high current peak. Therefore, a large PVR is expected to be achieved in these junctions.

\section{Model and calculation}

\begin{figure}[h]
    \begin{center}
     \includegraphics[angle = 0, width = 0.4\textwidth]{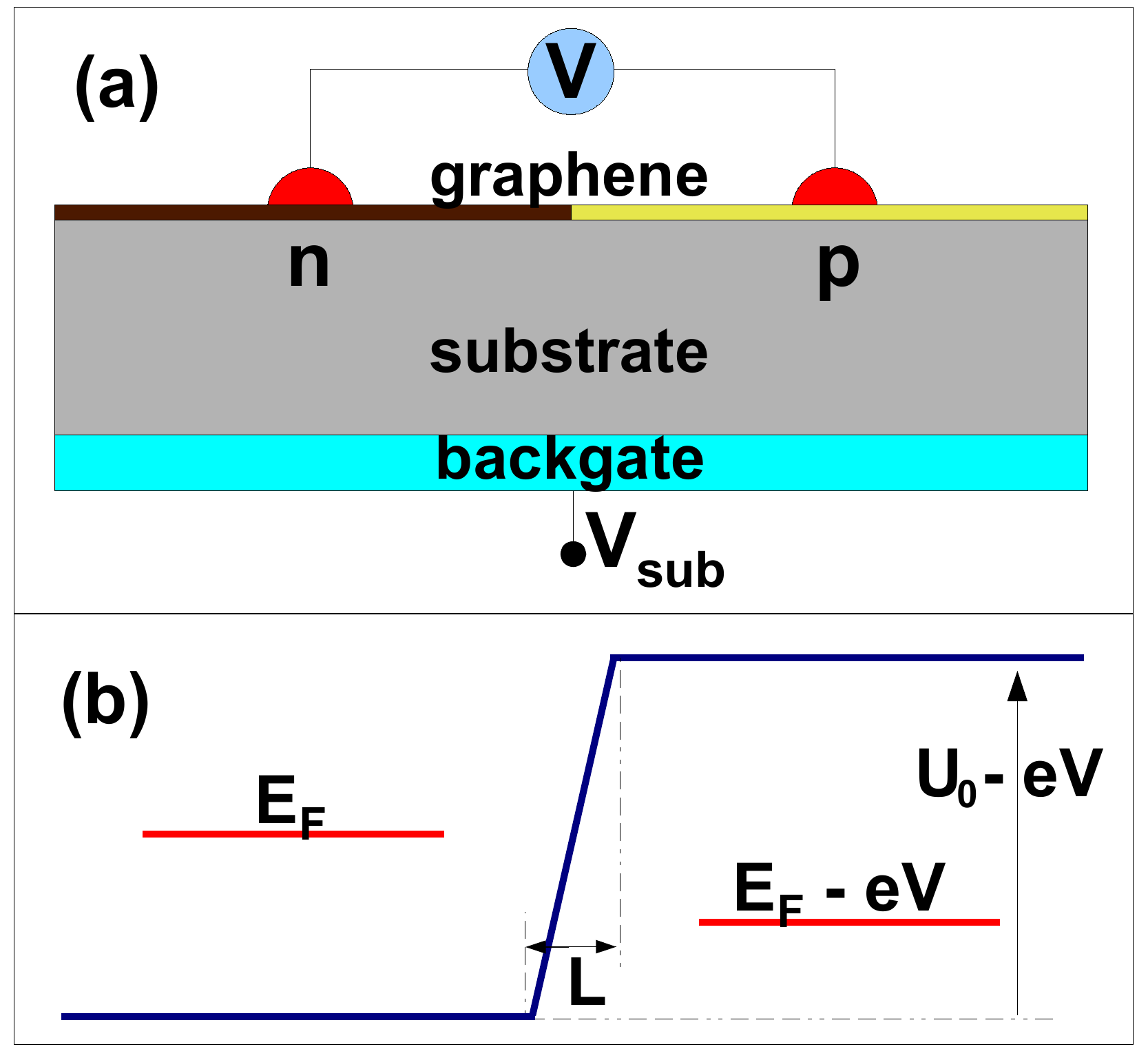}
    \end{center}
\caption{(Color online) Schematic of p-n junction based on monolayer graphene (a) and its potential profile along the transport direction (b). The n-doped (left) and the p-doped (right) regions are used to form a potential barrier in the device. The transition between two regions is characterized by the length $L$.}
\label{fig:num01}
\end{figure}
Graphene has a honeycomb lattice structure with a unit cell consisting of two carbon atoms - normally referred to as $A$ and $B$ atoms. To describe the charge states in the system, a simple nearest neighbor tight binding model can be conveniently used \cite{neto09}, with $a_c = 0.142$ nm as carbon-carbon distance, $t = 2.7$ eV as hopping energy between nearest neighbor sites, and $\varepsilon_A = - \varepsilon_B = \Delta$ as the onsite energies in the two sublattices. When $\Delta = 0$, this model results in a band structure with zero gap, i.e., the conduction and valence bands meet at the \textbf{K} and \textbf{K'} points in the Brillouin zone. By making $\Delta \neq 0$, the inversion symmetry is broken and therefore a finite gap opens in graphene's band-structure. The energy dispersion close to the \textbf{K}-point simply writes
\begin{equation}\label{eq01}
   E\left( {\vec k} \right) =  \pm \sqrt {\hbar ^2 v_F^2 \left( {k_x^2  + k_y^2 } \right) + \Delta ^2 }
\end{equation}
where $v_F = 3a_ct/2\hbar \approx 10^6$ m/s is the Fermi velocity, $\vec k = \left( {k_x ,k_y } \right)$ is the 2D-momentum, and the sign +/- stands for the conduction/valence band, respectively. From eq. (\ref{eq01}), the bandgap is determined as $E_g = 2\Delta$. To describe the excited states around such \textbf{K}-point, one can use the following effective (massive Dirac-like) Hamiltonian:
\begin{equation}\label{eq02}
   H =  - i\hbar v_F \left( {\sigma _x \partial _x  + \sigma _y \partial _y } \right) + \Delta \sigma _z  + U
\end{equation}
where $U$ is the external potential energy and $\sigma_{x,y,z}$ are the Pauli matrices. The Hamiltonian (\ref{eq02}) is now used to study the transport characteristics of the p-n junction schematized in Fig. 1. The p-doped and n-doped graphene regions can be generated by electrostatic doping \cite{huar07,zhan08} or by chemical doping \cite{farm09,bren10}. The back-gate is used to control the carrier density in the graphene layer by applying a voltage $V_{sub}$. The junction is characterized by the potential barrier $U_0$ and the transition length $L$. This length is the size of the region across which the charge density changes monotonically from n-type to p-type. Though expected to be short \cite{farm09}, it is finite and is considered in this work as a parameter.

In principle, the presence of defects and impurities is unavoidable and may result in a substantial amount of disorder in the graphene. However, such disorder affects strongly the electronic characteristics only when the graphene sheets are narrow \cite{cres08}. Thus, we assume that it can be negligible in this work where we consider the transport in wide systems \cite{vozm05}. Our study addresses the ballistic transport through the junction and the non-equilibrium Green's function (NEGF) technique (see the formulation in ref. \cite{vndo08}) is then used to solve eq. (\ref{eq02}) and to determine the transport quantities.
\begin{figure}[h]
    \begin{center}
     \includegraphics[angle = 0, width = 0.48\textwidth]{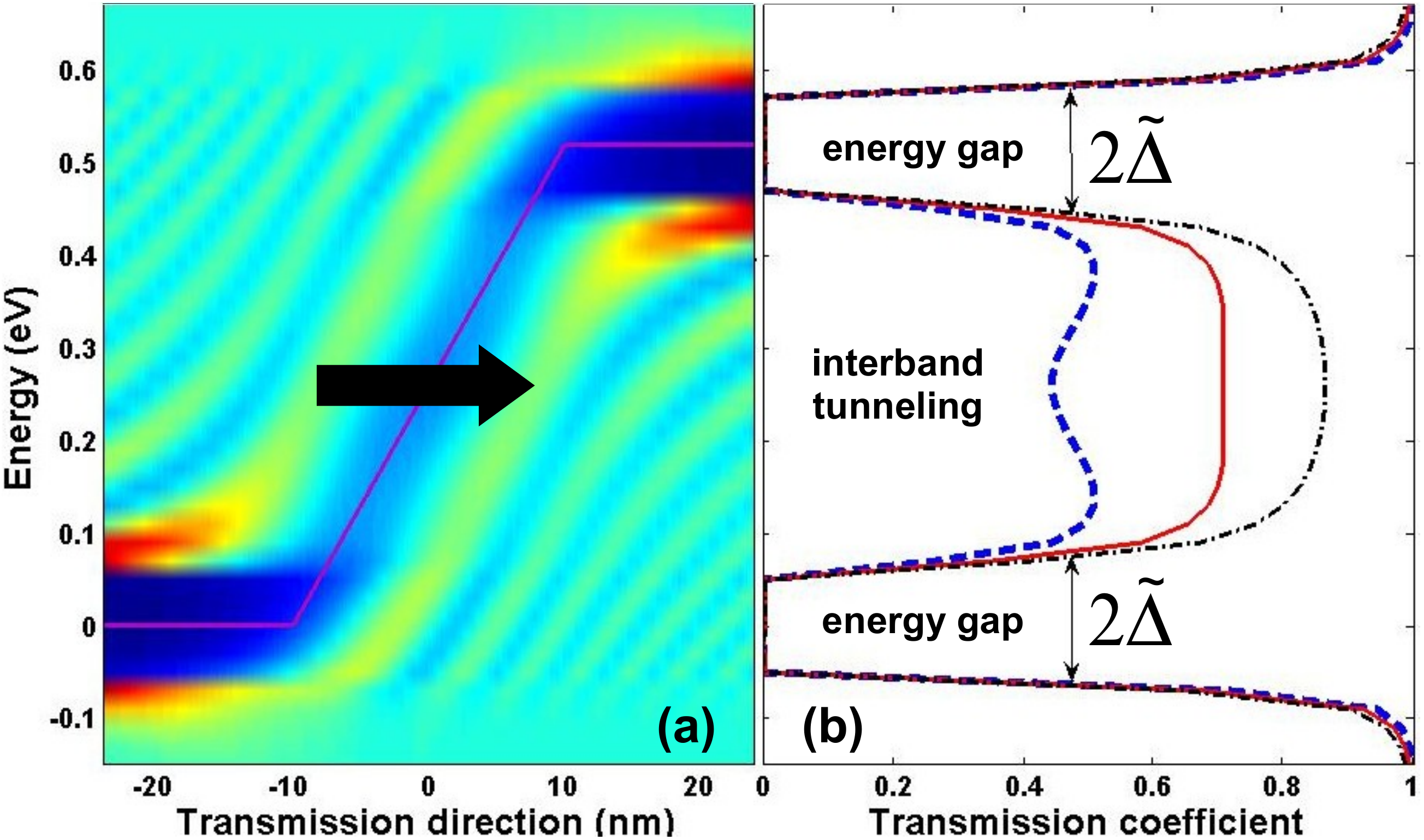}
    \end{center}
\caption{(Color online) Local density of states (a) and transmission coefficient (b) in a graphene p-n junction. The transition length in (a) $L = 20$ nm; in (b): $L = 5$ nm (dotted-dashed), 10 nm (solid) and 20 nm (dashed lines). All results are for $\tilde \Delta = 65$ meV ($\tilde \Delta  = \sqrt {E_y^2  + \Delta ^2 }$ and $E_y = \hbar v_F k_y$).}
\label{fig:num02}
\end{figure}

\section{Results and discussion}

\begin{figure}[h]
    \begin{center}
     \includegraphics[angle = 0, width = 0.48\textwidth]{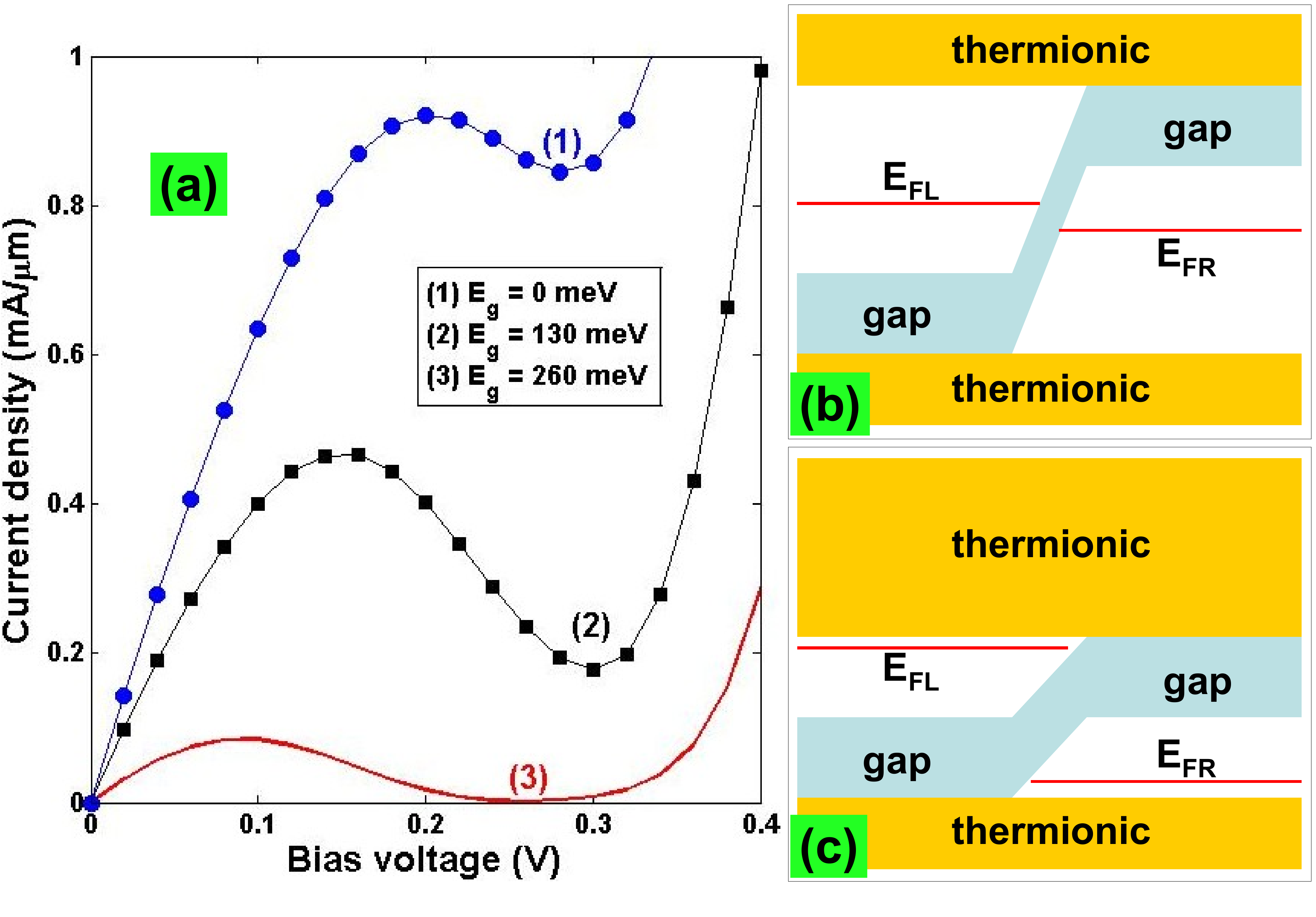}
    \end{center}
\caption{(Color online) (a) $I-V$ characteristics of graphene p-n junction with different energy bandgaps. Other parameters are: $E_F = 0.26$ eV, $U_0 = 0.52$ eV, and $L = 10$ nm. (b) and (c) illustrate the band diagrams at low bias and in the current valley, respectively.}
\label{fig:num03}
\end{figure}
Before investigating the behavior of the electrical current through these junctions, we plot a map of the local density of states and the transmission coefficient as a function of energy in Fig. 2(a) and Fig. 2(b), respectively. The results show clearly three important transport regions: (i) thermionic when $E < U_N - \tilde \Delta$ or $E > U_P + \tilde \Delta$, (ii) interband tunneling when $U_N  + \tilde \Delta  < E < U_P  - \tilde \Delta$, and (iii) transmission gap when $\left| {E - U_\alpha} \right| \leq \tilde \Delta$, where $U_\alpha \equiv U_{N,P}$ denote the potential energies in the n-doped and p-doped regions, respectively. The appearance of transmission gap is essentially due to the fact that the longitudinal momentum $k_x$ defined from eq. (\ref{eq01}) is imaginary in such energy region, i.e., the carrier states are evanescent and therefore decay rapidly when going to one of the two junction sides. Note that the same results are obtained for $\tilde \Delta = {\rm{const}}$ even with different $E_y$ and/or different $\Delta$. When $\Delta = 0$, the transmission gap is just a function of $E_y$ and disappears for normal incident particles. By rising $\Delta$, this gap increases and gets its minimum value $E_g = 2\Delta$ when $E_y = 0$. Moreover, due to the appearance of evanescent states around the neutral points in the transition region, Fig. 2(b) shows clearly that the larger the transition length, the lower the interband tunneling. However, it is worth noting that the interband tunneling of chiral fermions observed here is still very high in comparison with typical values of the order of $10^{-7}$ observed in conventional Esaki tunnel diodes \cite{carl97}. Therefore, a high current peak and then a large PVR are expected to be observed in the considered graphene junctions.

Now, in Fig. 3(a), we display the $I-V$ characteristics obtained for different energy bandgaps. Note that, throughout the work, the current is computed at room temperature. It is shown that the NDC behavior appears more clearly and its PVR increases when increasing the bandgap though the current peak reduces. The form of the $I-V$ curve can be explained well by looking at the diagrams in Fig. 3(b,c): at low bias (3(b)), the contribution of interband tunneling processes makes the current rising; when further rising the bias (3(c)), the interband tunneling is suppressed due to the transmission gaps and therefore the current decreases; when the bias is high enough, the contribution of charge carriers in the thermionic region makes the current rising very rapidly. Because the transmission gap increases, the current, especially in the valley region, decreases when increasing the bandgap. This results in a large PVR of NDC, for instance, about 46 in the case of $E_g = 260$ meV shown in Fig. 3(a).
\begin{figure}[h]
    \begin{center}
     \includegraphics[angle = 0, width = 0.43\textwidth]{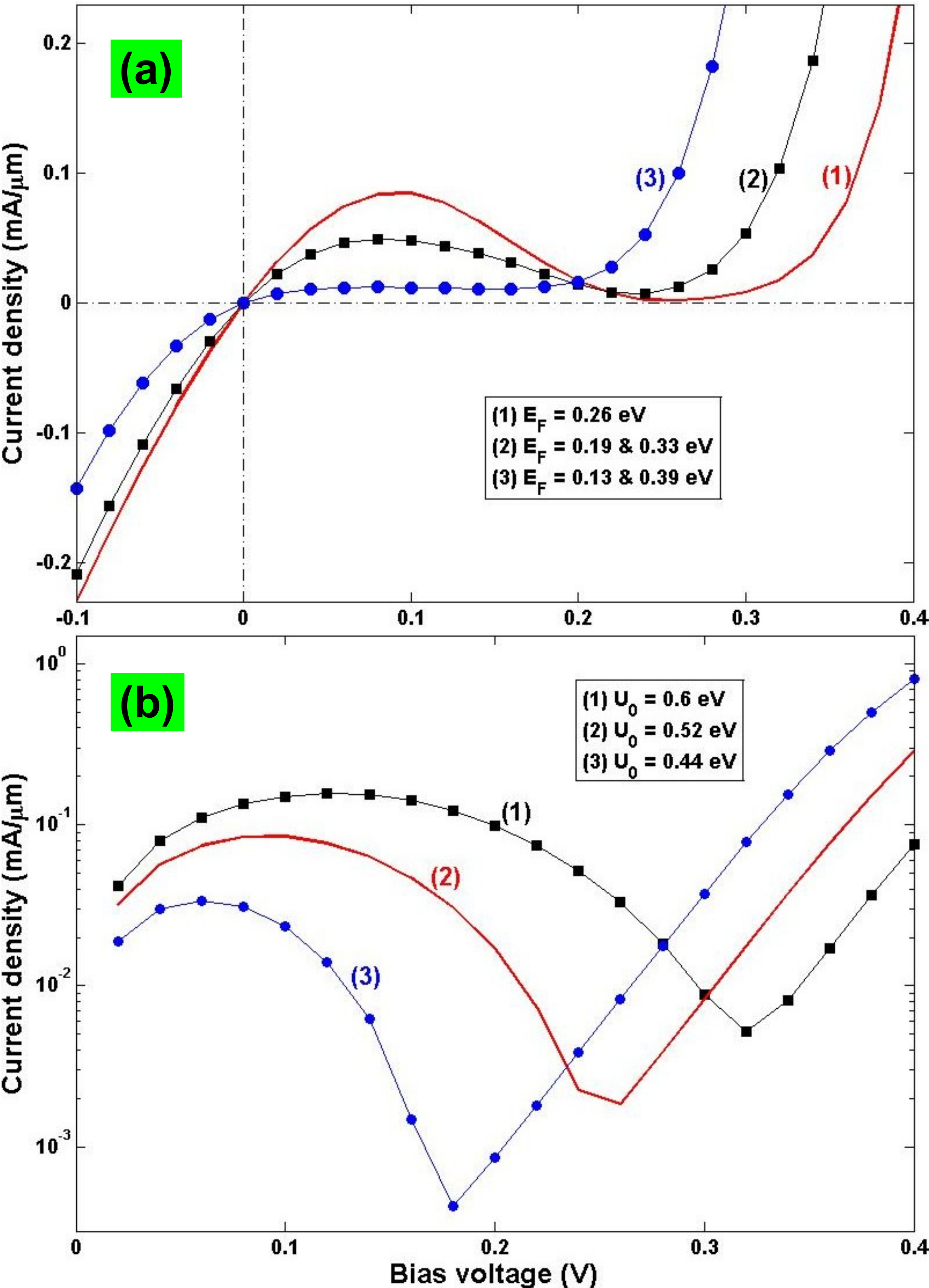}
    \end{center}
\caption{(Color online) $I-V$ characteristics of graphene p-n junction with different Fermi energies (a) and different potential barriers (b). Other parameters are: $E_g = 260$ meV, $E_F = U_N + U_0/2$ in (b), $U_0 = 0.52$ eV in (a), and $L = 10$ nm, where $U_N$ is the potential energy in the n-doped graphene.}
\label{fig:num04}
\end{figure}

\begin{figure}[h]
    \begin{center}
     \includegraphics[angle = 0, width = 0.48\textwidth]{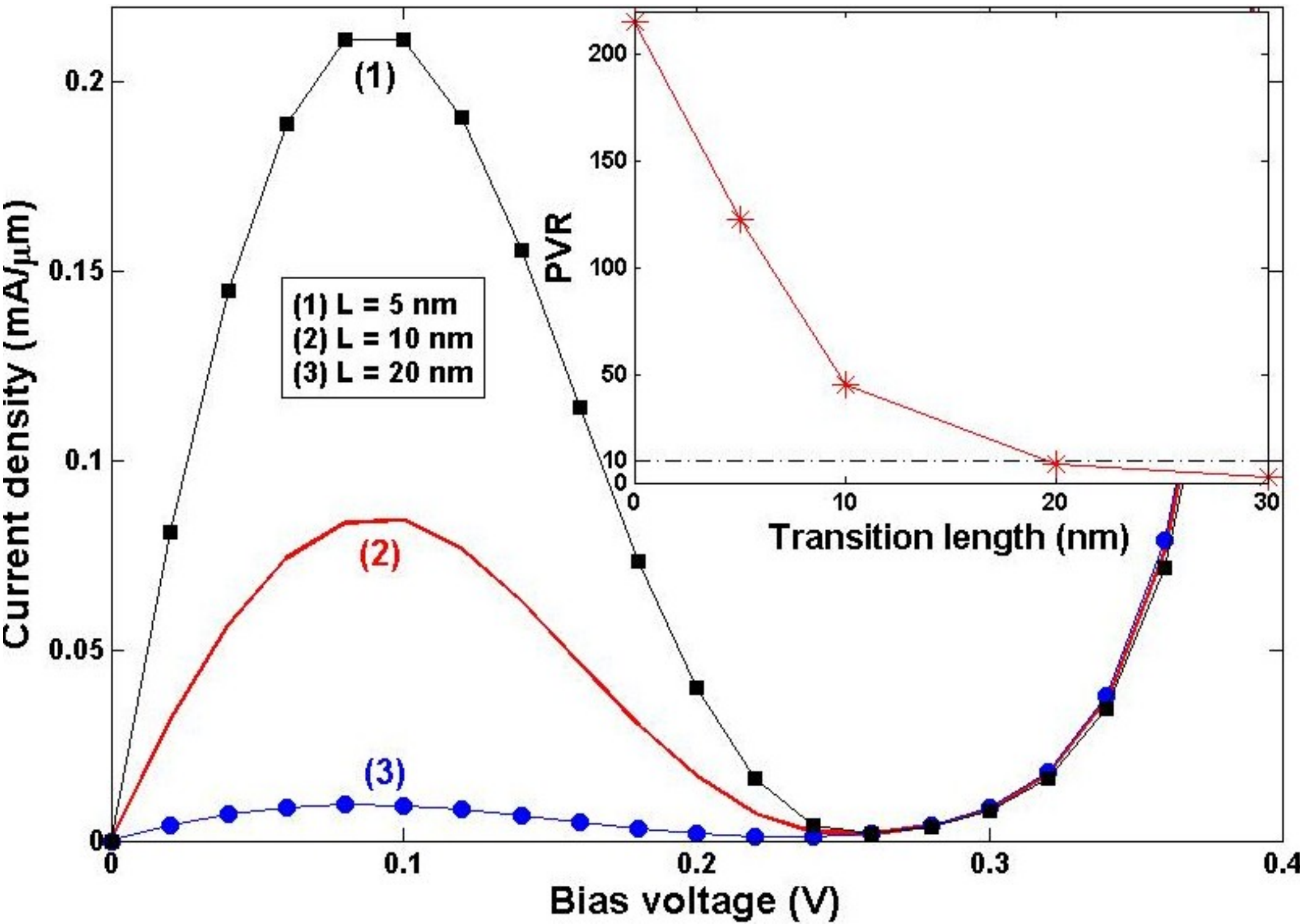}
    \end{center}
\caption{(Color online) $I-V$ characteristics of graphene p-n junction with different transition lengths. The inset shows the evolution of peak-to-valley ratio with respect to $L$. Other structure parameters are: $E_g = 260$ meV, $E_F = 0.26$ eV, and $U_0 = 0.52$ eV.}
\label{fig:num05}
\end{figure}
To evaluate the possibilities of having a strong NDC behavior in the junctions, we consider the behavior of the $I-V$ characteristics with different Fermi energies in Fig. 4(a) and different barrier heights in Fig. 4(b). Fig. 4(a) shows that the current valley shifts to the low bias region while the current peak decreases when $E_F$ goes from the value of $U_m = U_N + U_0/2$ to the top of the valence band in the p-doped region or to the bottom of conduction band in the n-doped one. Simultaneously, the current is a symmetric function of $E_F$ around $U_m$. This result leads to the fact that the strongest NDC is achieved when $E_F = U_m$ and the rectification behavior around the zero bias can be observed when $E_F$ goes away from such value, e.g., see the cases of $E_F = 0.13$ or 0.39 eV in Fig. 4(a). The latter behavior is essentially due to the role of the transmission gap (as illustrated in the diagrams in Figs. 3(b,c)), which induces a strong reduction of current in the low-positive bias region. Besides, when decreasing the barrier height, we find that the transmission gap around $U_P$ moves downward in energy, and as a consequence, the interband tunneling and the current peak are reduced. Indeed, this point is clearly illustrated in Fig. 4(b), where $E_F$ is chosen to be $U_N + U_0/2$ to achieve the strongest NDC. Moreover, it is shown that the current valley shifts to the low bias and its value is reduced, which finally results in an increase of the PVR when decreasing $U_0$. For instance, the PVR is $\sim 30$, $46$, and $79$ for $U_0 = 0.6$, $0.52$, and $0.44$ eV, respectively. To obtain the best device operation (either a high current peak or a large PVR), the study suggests that the barrier height about two times greater than the energy bandgap should be used.

Finally, the $I-V$ characteristics in the junctions with different transition lengths are displayed in Fig. 5. It is shown that because of the suppression of interband tunneling (as in Fig. 2(b)), the current peak is reduced when increasing $L$. Therefore, as seen in the inset of Fig. 5, the PVR decays as a function of the length $L$. However, the PVR as large as about 10 is still obtained even for $L$ up to 20 nm. This suggests the possibility of achieving a very large PVR with a short transition length, which can be realized by controlling the device geometry, e.g., by appropriately reducing the gate dielectric thickness in the case of the electrostatic doping \cite{zhan08}, or by using the chemical doping to generate the junction as mentioned in ref. \cite{farm09}. As expected above the large PVR, e.g., $\sim$ 123 obtained for $L = 5$ nm in this work, is much higher than for conventional Esaki tunnel diodes where its highest reported value is just about 16 (see in ref. \cite{oehm10} and references therein).

\section{Conclusion}
Using the NEGF technique, we have investigated the transport characteristics of monolayer graphene p-n junctions. Even at room temperature a negative-differential-conductance with peak-to-valley ratio as large as a few tens can be observed in these junctions thanks to a finite bandgap and to the high interband tunneling of chiral fermions. The dependence of this behavior on the device parameters was analyzed. It is shown that the strong negative-differential-conductance behavior is achieved when the Fermi level is in the center of the potential barrier, the barrier height is about two times greater than the bandgap, and the transition length is short. We hope that obtained results can be helpful for designing efficient room-temperature negative-differential-conductance devices based on finite-bandgap graphene.

\section*{Acknowledgments}
This work was partially supported by the French ANR through the projects NANOSIM-GRAPHENE (ANR-09-NANO-016) and MIGRAQUEL (ANR-10-BLAN-0304). One of the authors (VHN) acknowledges the Vietnam's National Foundation for Science and Technology Development (NAFOSTED) for financial support under the Project No. 103.02.76.09.

\end{document}